\documentclass{elsart}
\usepackage{epsfig}

\newcommand{\pT}{p_\perp}
\newcommand{\kTaveSq}{\mbox{$\left\langle k_\perp^2\right\rangle$}}

\newcommand{\Pythia}{\textsc{Pythia}}
\newcommand{\Jetset}{\textsc{Jetset}}
\newcommand{\Fritiof}{\textsc{Fritiof}}

\newcommand{\UNIT}[1]{\mbox{$\,{\rm #1}$}}

\newcommand{\GeV}{\UNIT{GeV}}
\newcommand{\GeVc}{\UNIT{GeV/c}}

\newcommand{\AGeV}{\UNIT{AGeV}}

\newcommand{\fm}{\UNIT{fm}}

\newcommand{\fmc}{\UNIT{fm/c}}
\newcommand{\proz}{\UNIT{\%}}

\newcommand{\SqrtS}[1]{\mbox{$\sqrt s=#1\GeV$}}

\newcommand{\PP}{p+p}
\newcommand{\PPbar}{p+$\bar{\rm p}$}

\begin{document}
\begin{frontmatter}
\title{Suppression of high transverse momentum hadrons at RHIC 
by (pre--) hadronic final state interactions}
\author[unig]{W.~Cassing},
\author[unig]{K.~Gallmeister\corauthref{cor1}}
\ead{Kai.Gallmeister@theo.physik.uni-giessen.de}
\author[unif]{C.~Greiner}
\corauth[cor1]{}
\address[unig]{Institut f\"ur Theoretische Physik, %
  Universit\"at Giessen, %
  Heinrich--Buff--Ring 16, %
  D--35392 Giessen, %
  Germany}
\address[unif]{Institut f\"ur Theoretische Physik, %
  Universit\"at Frankfurt, %
  Robert-Mayer-Str.~8--10, %
  D--60054 Frankfurt, %
  Germany}

\begin{abstract}
  We investigate transverse hadron spectra from proton+proton,
  deuteron+Au and Au+Au collisions at \SqrtS{200} and \SqrtS{17.3}
  within the Hadron--String--Dynamics (HSD) approach which is based on
  quark, diquark, string and hadronic degrees of freedom as well as
  \Pythia{} calculations for the high $\pT$ spectra.
  The comparison to experimental data on transverse mass spectra from
  p+p, d+Au and Au+Au reactions shows that pre--hadronic effects are
  responsible for both the hardening of the spectra for low transverse
  momenta (Cronin effect) as well as the suppression of high $\pT$
  hadrons. The interactions of formed hadrons are
  found to be negligible in central Au+Au collisions at
  \SqrtS{200} for $\pT\geq6\GeVc$, but have some importance
  for the shape of the ratio $R_{AA}$ at lower $\pT$ values
  ($\leq6\GeVc$). The large suppression seen
  experimentally is attributed to the inelastic interactions of 'leading'
  pre-hadrons with the dense environment, which could be partly
  of colored partonic nature in order to explain the large attenuation
  seen in most central Au+Au collisions.
\end{abstract}


\end{frontmatter}


\section{Introduction}

The phase transition from partonic degrees of freedom (quarks and
gluons) to interacting hadrons -- as expected in
the early universe shortly after the "big bang" -- is a central topic
of modern high--energy physics.
In order to understand the dynamics and relevant scales of
this transition laboratory experiments under controlled conditions
are presently performed with ultra--relativistic
nucleus--nucleus collisions.
Hadronic spectra and relative hadron abundancies from these
experiments reflect important aspects of the dynamics in the hot and
dense zone formed in the early phase of the reaction.
In fact, estimates based on the Bjorken formula \cite{bjorken} for the
energy density achieved in central Au+Au collisions suggest that the
critical energy density for the formation of a quark--gluon plasma
(QGP) of 0.7 to 1 GeV/fm$^3$ \cite{Karsch} is by far exceeded in the
initial phase for a
couple of $\fmc$ at Relativistic Heavy Ion Collider (RHIC) energies
\cite{QM01}, but it is still the aim to unambiguously identify the
formation and properties of this new phase.

Presently, transverse momentum (or transverse mass) spectra of
hadrons are in the center of interest. Here the suppression of
high transverse momentum hadrons is investigated in Au+Au
reactions relative to p+p collisions at RHIC energies of
\SqrtS{200} \cite{survey}, since the propagation of a fast quark
through a colored medium (QGP) is expected to be different from
that in cold nuclear matter as well as in the QCD vacuum due to
the energy loss of induced gluon radiation
\cite{WA1,WA2,Wang,Baier,vitev}. In fact, the PHENIX
\cite{PHENIX1}, STAR \cite{STAR1} and BRAHMS \cite{BRAHMS}
collaborations have reported a large relative suppression of
hadron spectra for transverse momenta above
$\pT\simeq3\cdots4\GeVc$ which might point toward the creation of
a QGP, since this suppression is not observed in d+Au interactions
at the same bombarding energy per nucleon
\cite{BRAHMS,PHENIX2,STAR2}. Accordingly, the experimental
observations are qualitatively in line with expectations from
perturbative Quantum Chromo Dynamics (pQCD) \cite{R1,R2,R3,R4,R5},
but do not support the idea of initial state gluon saturation
\cite{R6}. But, at present, it cannot be ruled out that a larger
fraction of the suppression seen in central Au+Au collisions might
be also due to hadronic final state interactions as suggested in
Ref.~\cite{Kai}. It is the aim of this work to quantify the amount
of hadronic final state interactions in d+Au and Au+Au collisions
at RHIC energies.

We employ the HSD transport model \cite{Ehehalt,Geiss,Cass99} for
our study. This approach takes into account the formation and
multiple rescattering of formed hadrons as well as unformed 'leading'
pre--hadrons and thus superseeds the
incoherent summation of individual p+p collisions. Such transport
calculations allow to study systematically the change in the
dynamics from elementary baryon--baryon or meson--baryon
collisions to proton--nucleus reactions or from peripheral to
central nucleus--nucleus collisions in a unique way without
changing any parameter. In the HSD approach nucleons, $\Delta$'s,
N$^*$(1440), N$^*$(1535), $\Lambda$, $\Sigma$ and $\Sigma^*$
hyperons, $\Xi$'s, $\Xi^*$'s and $\Omega$'s as well as their
antiparticles are included on the baryonic side whereas the $0^-$
and $1^-$ octet states are included in the mesonic sector. Inelastic
hadron--hadron collisions with energies above $\sqrt s\simeq 2.6\GeV$
are described by the \Fritiof{} model \cite{LUND} (including
\Pythia{} v5.5 with \Jetset{} v7.3 for the production and
fragmentation of jets \cite{PYTHIA0}) whereas
low energy hadron--hadron collisions are modeled in line with
experimental cross sections. We mention that no explicit parton
cascading is involved in our transport calculations contrary to
e.g.~the AMPT model \cite{Ko_AMPT}.

A systematic analysis of HSD results and
experimental data for central nucleus--nucleus collisions for
$2\cdots160\AGeV$ has shown that the spectra for the 'longitudinal'
rapidity distribution of protons, pions, kaons, antikaons and
hyperons are in reasonable agreement with available data. Only the
pion rapidity spectra are slightly overestimated from AGS to SPS
energies \cite{Weber02} which implies, that the maximum in the
$K^+/\pi^+$ ratio at $20\cdots30\AGeV$ -- seen in central Au+Au
(Pb+Pb) collisions \cite{NA49} -- is missed.
For a comparison of HSD calculations with experimental data at RHIC
energies we refer the reader to Ref.~\cite{Brat03}.

In this work we concentrate on the transverse momentum dynamics
and especially on the very high momentum tail of the hadron
spectra. In order to describe these high $\pT$ spectra, we use the
\Pythia{} v6.2 event generator \cite{PYTHIA} for nucleon--nucleon
collisions, which describes the high transverse momentum spectra
of peripheral nucleus--nucleus collisions from RHIC at \SqrtS{130}
very well \cite{Kai}. We recall, that within this framework also
experimental dilepton and direct photon spectra at SPS energies
\cite{KaiSPSLepton} or single lepton spectra at RHIC energies from
open heavy quark production \cite{KaiRHICLepton} are well
reproduced.

\section{Transport calculations versus experimental data}

\begin{figure}[htb!]
  \begin{center}
    \includegraphics[width=8cm]{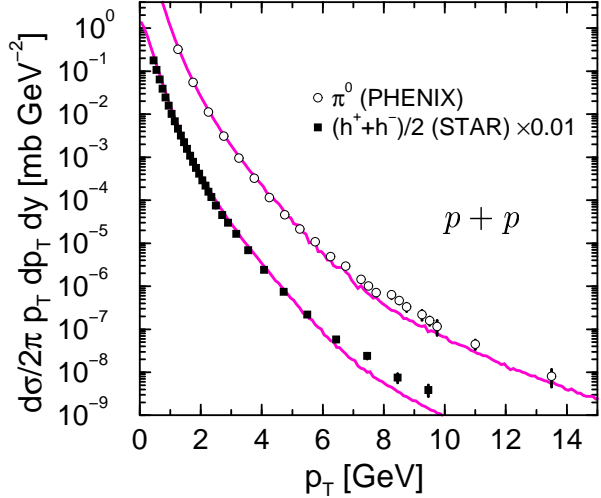}

    \caption{
      The invariant cross sections for the production of neutral pions
      and charged hadrons in p+p collisions (\SqrtS{200}) at
      midrapidity as a function of transverse momentum.
      The \Pythia{} (version v6.2) calculations (solid lines) are compared to the
      experimental data from RHIC  (open circles: PHENIX
      \cite{PHENIX_pp_pi0}, filled squares: STAR \cite{STAR1}).
      The charged hadron spectra are divided by a factor of 100 for clarity.
      }
    \label{fig1}
  \end{center}
\end{figure}

We start with hadron production in p+p collisions at the invariant
energy \SqrtS{200}.  The comparison of the calculations
for the transverse momentum spectra of charged hadrons from p+p
collisions at midrapidity with the experimental data of the
PHENIX and STAR collaborations \cite{PHENIX_pp_pi0,STAR1} is
shown in Fig.~\ref{fig1}, which demonstrates that the description
of hadron production in the elementary reaction is quite well
under control (see also Ref.~\cite{Kai}).

\subsection{Preliminaries}

\begin{figure}[htb!]
  \begin{center}
    \includegraphics[width=8cm]{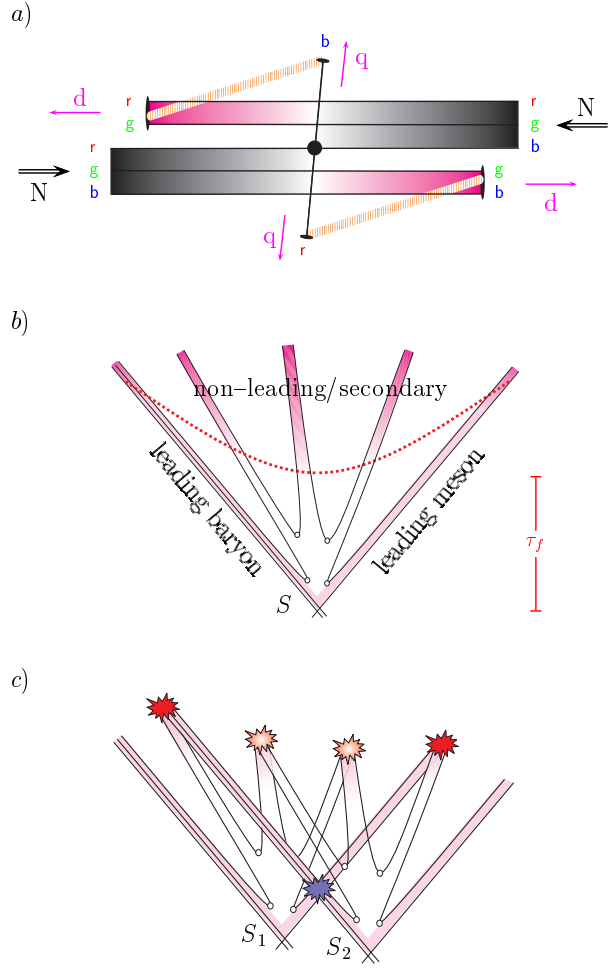}
    \caption{%
      Upper Part: Illustration of the formation of two strings in a
      N+N collision. The formed strings are color--neutral in total,
      however, carry (anti--)color at the string--ends, which here are
      displayed as diquark ($d$) -- quark ($q$) pairs.
      The hatched areas denote nonperturbatively interacting
      partons. 
      Middle Part: The space--time evolution of a string $S$ in its rest
      frame. Whereas the 'leading' (di--)quarks
      (represented by the outer lines) form 'leading' pre--hadrons
      almost instantly, secondary $q\bar q$ pairs are excited in the
      non--perturbative vacuum (after some time delay $t_1$) and
      recombine to secondary hadrons (shadowed double lines).
      The time from the initial creation of the string to the final
      formation of secondary hadrons is denoted as $\tau_f$. The
      distribution in the formation time is presently unknown and for
      convenience characterized by a single formation time $\tau_f$ for all
      hadrons (dotted hyperbola). 
      Lower Part: Whereas 'secondary hadrons' are not allowed to
      scatter during their formation time, they interact with the
      full hadronic cross section after being formed. The 'leading'
      color neutral pre--hadrons, however, carrying a constituent
      quark from the struck hadron, may scatter (in--)elastically with
      other leading pre--hadrons as well as formed secondary hadrons
      with a reduced cross section.
      In the present realization of HSD this reduced cross section is
      determined in line with the constituent quark model (see text).
      }
    \label{fig2}
  \end{center}
\end{figure}

Before coming to proton+nucleus or deuteron+nucleus collisions it
is important to specify the concept of ``leading'' (pre--)hadrons and
``non--leading'' or ''secondary'' hadrons since this separation is
of central importance for the results to be discussed below. To
this aim we recall that in the string picture -- in a high energy
nucleon+nucleon collision -- two (or more) color--neutral strings
are assumed to be formed, which phenomenologically describe the
low energy ``coherent'' gluon dynamics by means of a color
electric field, which is stretched between the 'colored' ends of each
single string (cf.~Fig.~\ref{fig2}a for an illustration).
The latter string ends are defined by the space--time coordinates of
the constituents, i.e.~a diquark and quark for a ``baryonic'' string
or quark and antiquark for a ``mesonic'' string. These constituent
quarks, diquarks or antiquarks are denoted as ``leading'' quarks.
In line with Ref. \cite{Kopel4} these constituent 'partons' are
assumed to pick up (almost instantly) an anti-colored parton from the
vacuum and  achieve color neutrality.
These color-neutral objects -- containing the string ends -- are
denoted as ``leading pre--hadron'' (cf.~Fig.~\ref{fig2}b for an
illustration of the string fragmentation in space and time in the rest
frame of the string.)

In principle, a leading pre--meson or pre--baryon is able to hold up
to 2 (3) leading quarks, but in the case of interest, i.e.~midrapidity
transverse momentum spectra in heavy--ion collision at \SqrtS{200},
the majority of leading mesons/hadrons consists only of
one leading quark and some secondary (di--) quarks, as will be
discussed below. We note, that this changes dramatically for rapidity
regions closer to those of the projectile or target or different
$\sqrt s$ values, as discussed below.

The time that is needed from the instant of the 'hard' pQCD
collision $t_0=0$ to the formation of $q\bar{q}$
or $qq\overline{qq}$ pairs from the vacuum ($t_1 > t_0$) and
for the hadronization of the fragments ($t_2=\tau_f$) we denote as
formation time $\tau_f$ in line with the convention in transport
models. For
simplicity we assume (in HSD and also for the high $\pT$ partons),
that the formation time is a constant $\tau_f$ in the rest frame of
each hadron and that it does not depend on the particle species.
Other estimates (cf.~\cite{Kai,URQMD1,URQMD2}) show a more complicated
structure and particle dependence (especially for the light pion), but
this issue will not be discussed in detail here; nevertheless, we will
show results for different formation times and alternative
scenarios below.
In principle one expects also a distribution in the formation times,
however, we here address only the mean value of such an 'unknown'
distribution. We recall, that due to time dilatation the formation
time $t_f$ in any reference frame is then proportional to the energy
of the particle
\begin{equation}
\label{eq:formation-time}
        t_f=\gamma\cdot\tau_f=\frac{E_h}{m_h}\cdot\tau_f\ .
\end{equation}
The size of $\tau_f$ can roughly be estimated by the time that the
constituents of the hadrons (with velocity $c$) need to travel a
transversal distance of a typical hadronic radius ($0.5\cdots0.8\fm$).
Furthermore, since after the formation of a color neutral
pre-hadron ($t \geq t_1)$ the Fock-components of this object are
widespread in mass according to the uncertainty relation in energy
and time, it will take a time of $\Delta E^{-1}$ to form a
specific hadronic state of well defined radial excitation, where $\Delta
E$ denotes the level distance of the hadronic spectrum.

We assume in our transport simulations that hadrons, whose constituent
quarks and antiquarks are created from the vacuum in the string
fragmentation (at times $t_1$), do not interact with the surrounding nuclear medium
within their formation time (cf.~Fig.~\ref{fig2}c for an illustration).
For the leading pre-hadrons, i.e.~those involving (anti--)quarks or
diquarks from the struck nucleons, we adopt a reduced effective cross
section $\sigma_{\rm lead}$ during the formation time $\tau_f$ and the
full hadronic cross section later on. Since this rough
approximation is subject of current debate \cite{Kopel4} we
 will also present alternative models for the cross section
of the leading pre-hadrons (see below). The 'default' concept is
illustrated in more detail in Fig.~\ref{fig2}c where a 'star' signals
a possible collision while crossing lines imply that no collision
might occur.

Due to time dilatation light particles emerging from the middle of
the string can escape the formed 'fireball' without further
interaction, if they carry a high momentum relative to the rest
frame of the fireball. We note that these arguments are strictly
valid only in case of a constant formation time $\tau_f$ in the rest
frame as described above and implemented here.
However, hadrons with transverse momenta larger than $6\GeVc$
predominantly stem from the string ends and therefore can interact
directly with a reduced cross section (see below).

\subsection{Numerical implementation}

For the production and propagation of hadrons with high transverse
momentum ($> 1.5\GeVc$) we employ a perturbative scheme as also used
in Refs.~\cite{Cass01,Geiss99,Cass97,CassKo} for the charm and open charm
degrees of freedom.
Each high $\pT$ hadron is produced in the transport calculation with
a weight $W_i$ given by the ratio of the actual production cross
section divided by the inelastic nucleon--nucleon cross section, e.g.
\begin{equation}
  W_i = \frac{\sigma_{NN \rightarrow h(\pT) +
      x}(\sqrt{s})}{\sigma_{NN}^{\rm inelas.}(\sqrt{s})}.
\end{equation}
In the transport simulation we follow the motion of the high $\pT$
hadrons within the full background of strings/hadrons by propagating
them as free particles, i.e.~neglecting in--medium potentials, but
compute their collisional history with baryons and mesons or quarks
and diquarks.
For reactions with diquarks we use the corresponding reaction cross
section with baryons multiplied by a factor of 2/3.
For collisions with quarks (antiquarks) we adopt half of the cross
section for collisions with mesons and for the leading pre-hadron (formed)
baryon collision a factor of 1/3 is assumed. The elastic and
inelastic interactions with their fractional cross section are
modeled in the HSD approach in the same way as for ordinary
hadrons with the same quantum numbers via the \Fritiof{} model \cite{LUND} (including
\Pythia{} v5.5 with \Jetset{} v7.3 for the production and
fragmentation of jets \cite{PYTHIA0}).

The relative quark counting factors mentioned above might appear
arbitrary and simplistic. However, this concept -- oriented along
the additive quark model -- has been proven to work rather well
for nucleus--nucleus collisions from AGS to RHIC energies
\cite{Weber02,Brat03,Brat03b} as well as in hadron formation and attenuation
in deep inelastic lepton scattering off nuclei \cite{Falter}.
Especially the latter reactions are important to understand the
attenuation of pre--hadrons or ordinary hadrons with high momentum
in cold nuclear matter \cite{Kopel4}. Our studies in Ref.~\cite{Falter} have
demonstrated that the dominant final state interactions (FSI) of
the hadrons with maximum momentum, as measured by the HERMES
collaboration \cite{HERMES}, are compatible to the concepts
described above. This also holds for antiproton production and
attenuation in proton--nucleus collisions at AGS energies
\cite{AGS02}. Both independent studies point towards a hadron
formation time $\tau_f$ in the order of $0.4\cdots0.8\fmc$.

\begin{figure}[htb!]
  \begin{center}
    \includegraphics[width=8cm]{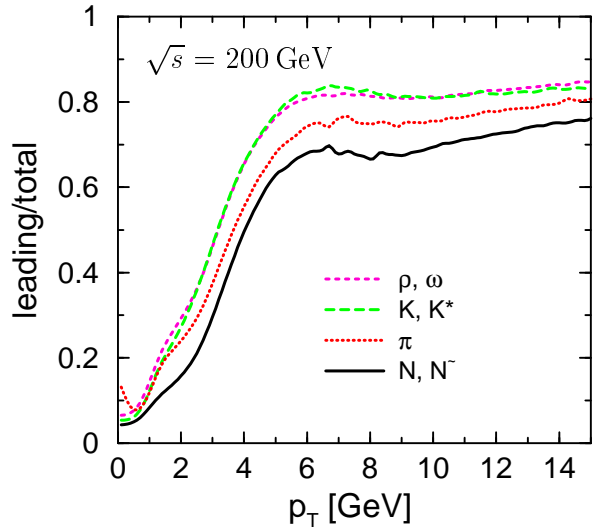}

    \caption{
      The ratio of 'leading particles' to 'all produced particles'
      (before particle decays) in N+N collisions (\SqrtS{200})
      at midrapidity for different particle classes as a function of
      transverse momentum within the \Pythia{} description.
      (An isospin average in line with central Au+Au collisions
      has been applied.)
      For the region $\pT=5\cdots7\GeVc$ the theoretical error is of
      the order $\pm0.05$ and to be neglected outside this region.
      }
    \label{fig3}
  \end{center}
\end{figure}

As mentioned above, a crucial question for the interaction dynamics is
the fraction of leading pre--hadrons to secondary hadrons (or all produced
hadrons) as a function of $\pT$ for the different hadron species.
This information is directly extracted from the \Pythia{} calculations
for different hadrons by explictly tracing back the string
fragmentation histories with all their details, i.e., beginning with the
incoming hadrons, following the hard interaction and linking the quark
content of the final hadrons back to the partons at the initial string ends.
The resulting fraction of leading pre--hadrons to all produced hadrons
is displayed in Fig.~\ref{fig3} for N+N collisions at \SqrtS{200}.
We have checked, that in Fig.~\ref{fig3} contributions with 2 (or
more) leading particles are negligible and thus the above
mentioned ``leading'' cross sections $\sigma_{\rm lead}$ as a
fraction of the hadronic cross section of 1/2 for mesons and 1/3
for baryons are sustained by simple quark counting rules.

In detail, one notices slight differences between pions and especially
antibaryons (not shown), however, the fraction of leading pre--hadrons
increases almost linearly with $\pT$ and approximately saturates above
$\pT\simeq6\GeVc$ for $\sqrt{s}=200\GeV$.
Thus at high momenta the major fraction of 'hadrons' is of ``leading''
origin, i.e.~pre--hadronic states containing quarks, antiquarks or
diquarks from the primary struck nucleons, and may interact according to the
assumptions stated above without delay with the (pre--) hadronic
environment (cf.~Fig.~\ref{fig2}c).

\subsection{Initial state Cronin enhancement}

As known from the experimental studies of
Refs.~\cite{Cronin1,Cronin2} an enhancement of the high transverse
momentum particle cross section from proton--nucleus collisions --
relative to scaled p+p collisions -- is already observed at SPS
and ISR energies. This 'Cronin effect' is presently not fully
understood in its details, but probably related to an increase of
the average transverse momentum squared $\kTaveSq$ of the partons
in the nuclear medium. One may speculate that such an enhancement
of $\kTaveSq$ is due to induced initial semi--hard gluon radiation
in the medium, which is not present in the vacuum due to the
constraint of color neutrality. A related interpretation has been
given by e.g.~Kopeliovich et al.~\cite{R3,Kopel4}. Since the
microscopic mechanisms are beyond the scope of our present
analysis, we do not want to comment this any further and simulate
this effect in the transport approach by increasing the average
$\kTaveSq$ in the string fragmentation with the number of previous
collisions $N_{\rm prev}$ as
\begin{equation}
  \label{kt}
   \kTaveSq= \kTaveSq_{pp} (1+\alpha N_{\rm prev})\quad.
\end{equation}
The parameter $\alpha \approx 0.25 - 0.4$ in (\ref{kt}) is fixed in
comparison to the  experimental data for d+Au collisions
\cite{PHENIX2,STAR2} (see below).
The assumption (\ref{kt}) is  in line with an independent suggestion in
Ref.~\cite{Papp}.

\begin{figure}[htb!]
  \begin{center}
    \includegraphics[width=8cm]{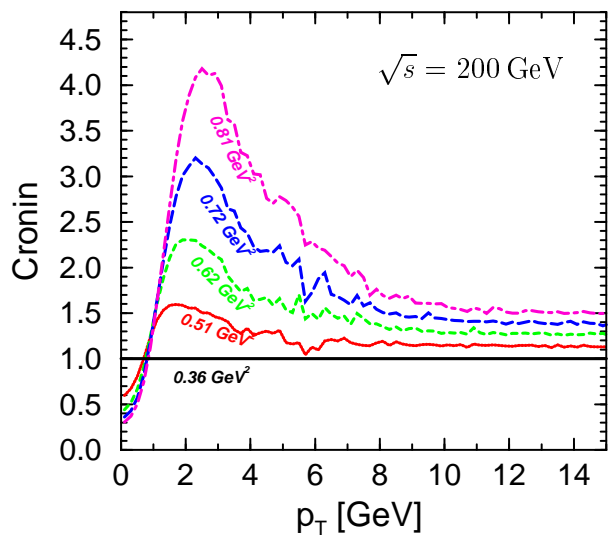}

    \caption{
      The Cronin enhancement factor (see text) as a function of the
      transverse momentum $\pT$ of charged hadrons for different
      average $\kTaveSq$ according to the \Pythia{} calculations
      for N+N collisions at \SqrtS{200}.
      }
    \label{fig4}
  \end{center}
\end{figure}

In order to show the effect of enhancing the average $\kTaveSq$
in the fragmentation of the string we show in Fig.~\ref{fig4} the
ratio of the charged hadron spectra from $pp$ collisions for
different $\kTaveSq$ to the spectra for the default value of
$\kTaveSq_{pp} = 0.36\GeV^2$.
These ratios may be denoted as 'Cronin enhancement factors', which are
a function of $\kTaveSq$ and the hadron species.
We note, that these 'Cronin enhancement factors' enter the dynamical
transport calculations for each hadron type separately, whereas only
the average enhancement factor is displayed for charged hadrons in
Fig.~\ref{fig4}.

The effect of such Cronin enhancement is obvious from Fig.~\ref{fig4}:
at low transverse momentum one observes a suppression of the $\pT$
spectrum with increasing $\kTaveSq$, which leads to a hardening of
hadron spectra from nucleus--nucleus collisions relative to p+p
reactions for transverse masses below about $(m_\perp-m_0)\leq
1\GeV$. A maximum in the enhancement factor arises at $\pT\sim 2\dots3\GeVc$
followed by a smooth decrease towards unity at high transverse
momenta, but a sizeable (constant) enhancement (above unity) remains
up to the highest momenta considered in our calculations.

\subsection{d+Au collisions}

\begin{figure}[htb!]
  \begin{center}
    \includegraphics[width=8cm]{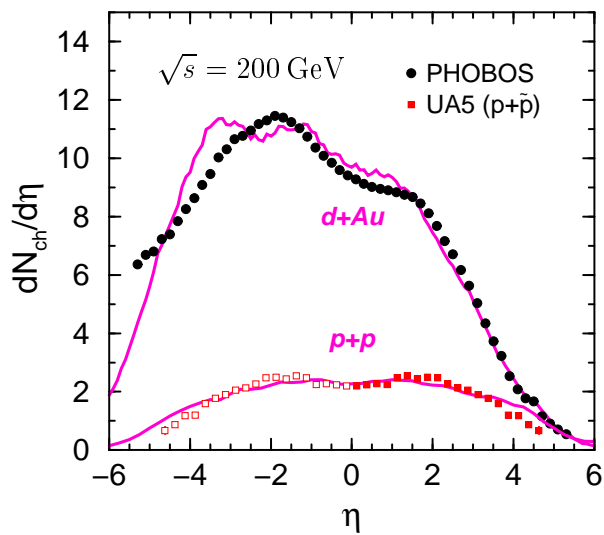}

    \caption{
      The calculated pseudorapidity distributions $dN/d\eta$
      of charged particles from inelastic p+p and d+Au
      reactions at \SqrtS{200} (solid lines) in comparison
      with the data for \PPbar{} from the UA5 Collaboration \cite{UA5}
      and d+Au collisions from the PHOBOS Collaboration
      \cite{PHOBOS03}.
      [The experimental systematic errors for  d+Au  are on the level of 20\proz{},
      whereas those in the normalization of our transport calculations
      are less than 10\proz{}.]
      }
    \label{fig5}
  \end{center}
\end{figure}

Whereas elementary p+p collisions give pseudorapidity distributiuons
$dN_x/d\eta$ of produced hadrons $x$ symmetric in $\eta$
(Fig.~\ref{fig5}), the recent measurements of 
$dN_x/d\eta$ of primary charged hadrons in minimum bias d+Au
collisions at \SqrtS{200} from the PHOBOS collaboration
\cite{PHOBOS03} show a significant asymmetry in $\eta$
(Fig.~\ref{fig5}). 
This asymmetry in $\eta$ is quite well reproduced by our calculations
for minimum bias d+Au collisions (Fig.~\ref{fig5}) as well as the
approximate maximum in $dN/d\eta$ for $\eta \approx -2$.
This implies that the rescattering of pre-hadrons from the
deuteron projectile ($\eta >0$) with nucleons of the target ($\eta
< 0$) is essential and that particles with large longitudinal
momentum ($\eta\geq 2$) are absorbed substantially in the
Au--target ending up at pseudorapidity ranges closer to the target
pseudorapidity.

In order to sustain these findings, Fig.~\ref{fig5} shows also
the comparison of our model calculations for \PP{} with data for
\PPbar{} from UA5 \cite{UA5} at the same invariant energy.
The difference between the data and the calculations is on the
10\proz{} level and partly due to the fact, that UA5 measures
inelastic events for \PPbar{}, whereas the calculations are for \PP{}
collisions:
We attribute the differences for pseudorapidity regions close to
those of the initial baryons (antibaryon) to an excess of light quarks
in the \PP{} reaction.

Thus, in addition to the comparison in Fig.~\ref{fig1} for the
transverse momentum distribution also the (soft) longitudinal hadron
production is reasonably well described in our transport approach for
elementary p+p as well as d+Au reactions.

Since the Cronin enhancement factor essentially depends on
$\kTaveSq$, which in turn is controlled by the parameter
$\alpha$ in eq.~(\ref{kt}), we will determine its range by the
data on d+Au reactions.

\begin{figure}[htb!]
  \begin{center}
    \includegraphics[width=8cm]{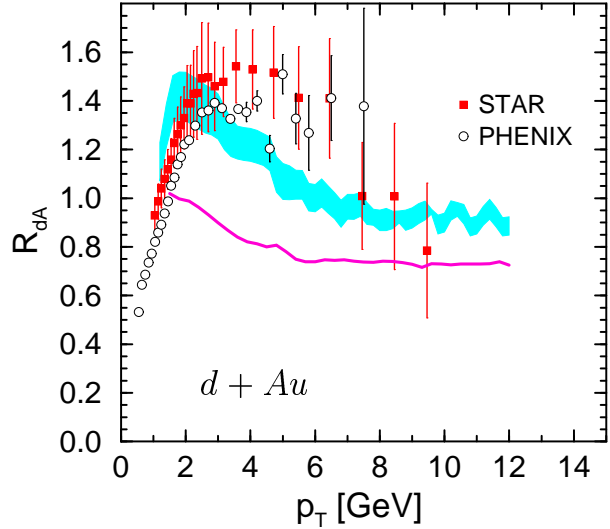}

    \caption{
      The suppression factor $R_{\rm dA}$ for minimum bias
      d+Au collisions (\SqrtS{200}) at midrapidity for
      charged hadrons.
      Experimental data are from PHENIX \cite{PHENIX2} (open circles)
      and STAR \cite{STAR2} (filled squares).
      The hatched band denotes our calculations for the 'Cronin'
      parameter $\alpha = 0.25\cdots0.4$ in (\ref{kt}) while
      the solid line results from transport calculations
      without employing any initial state Cronin enhancement ($\alpha=0$).
      }
    \label{fig6}
  \end{center}
\end{figure}

A comparison of the calculated ratio
\begin{equation}
  \label{ratio}
  R_{\rm dA}(\pT) = \frac{1/N_{\rm dA}^{\rm event}\ d^2N_{\rm dA}/dy d\pT}
  {\left<N_{\rm coll}\right>/\sigma_{pp}^{\rm inelas}\ d^2 \sigma_{pp}/dy d\pT}
\end{equation}
is shown in Fig.~\ref{fig6} with the respective data for charged
hadrons from the PHENIX and STAR collaborations
\cite{PHENIX2,STAR2} using $\alpha= 0.25\cdots0.4$ (hatched band).
In eq.~(\ref{ratio}) $\langle N_{\rm coll}\rangle \approx 8.5$ denotes
the number of inelastic nucleon--nucleon collisions per event (for
minimum bias collisions), whereas $\sigma_{pp}^{\rm inelas}$
is the inelastic p+p cross section\footnote{We remark  with some caution,
that the underlying binary
  scaling according to the Glauber model has recently been
  questioned by Kopeliovich \cite{Kopel3} since inelastic shadowing
  corrections should reduce the total inelastic cross section by
  up to 20\%.}.
We find that our calculations give a rise in the
ratio eq.~(\ref{ratio}) for small $\pT$ and a slight decrease
for $\pT\geq 2.5\GeVc$ which can be traced back to inelastic
interactions of the leading pre-hadrons with the nucleons in the
target. The description of the data for $R_{\rm dA}(\pT)$ is not
perfect and the data even indicate a slightly larger enhancement
for $\pT=4\dots6\GeVc$, but the agreement is
sufficient to proceed with Au+Au collisions at the same
$\sqrt{s}$. The important issue to remember in this context is that no
dramatic absorption of high $\pT$ hadrons in $d+Au$ collisions is
found in the calculations as well as in the data.

We note that the number of inelastic collisions
$\langle N_{coll}\rangle$ entering the ratio eq.~(\ref{ratio}) are
directly available in the transport calculations; thus no model dependence
enters the computation of this ratio.

For completeness we show in Fig.~\ref{fig5} also the result for
the ratio (\ref{ratio}) from a calculation without any initial
state Cronin enhancement (solid line). In this limit the ratio
drops below unity due to the interactions of the leading
pre-hadrons with the residual target nucleons.

\subsection{Au+Au collisions}
As in case of the d+Au system we expect also an attenuation of
high momentum hadrons in Au+Au reactions due to inelastic
interactions of leading pre-hadrons with the nucleons from the
target (projectile). In addition, inelastic interactions between
the pre-hadrons among each other should occur as well as between
leading pre-hadrons and formed secondary hadrons. Also
interactions of formed secondary hadrons between each other might
contribute to the final attenuation of high $p_T$ particles, if
they are formed sufficiently early inside the hot/dense fireball
\cite{Kai}. We mention, that in the calculations to be shown below
there are no interactions between leading pre-hadrons and explicit
quark or antiquarks or gluons in the very early phase of the
collision.

Furthermore, in the HSD approach the formation of secondary hadrons is
not only controlled by the formation time $\tau_f$, but also by
the energy density in the local rest frame, i.e., hadrons are not
allowed to be formed if the energy density is above 1 GeV/fm$^3$
\cite{Weber02}. This energy density cut in HSD prevents hadrons to be
formed in central nucleus-nucleus collisions at RHIC for a couple of fm/c
when setting the clock by the initial impact of the two ions. We
note in passing, that a reduction of this energy density cut to
0.7 GeV/fm$^3$ (cf. Introduction) leads to a further delay of secondary hadron
formation by $\sim$ 1.2 fm/c. Recall, however, that this energy cut
does not apply for the pre-hadronic states defined above.

\begin{figure}[htb!]
  \begin{center}
    \includegraphics[width=8cm]{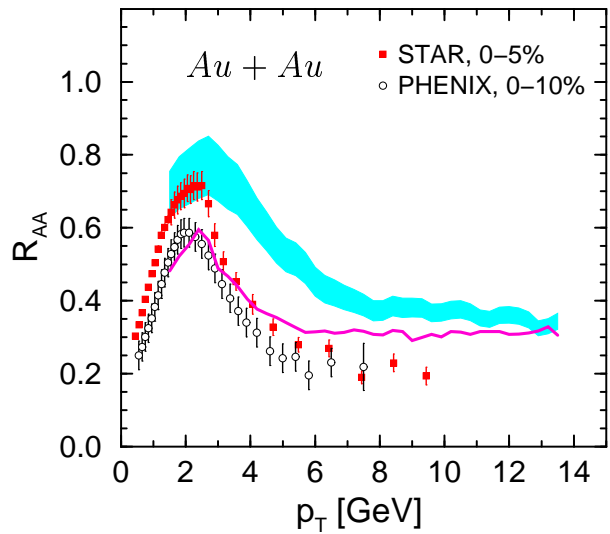}

    \caption{
      The suppression factor $R_{\rm AA}$ of charged hadrons
      at $0\cdots5\proz{}$ central Au+Au collisions (\SqrtS{200})
      at midrapidity;
      Experimental data are from
      Refs.~\cite{PHENIX1,STAR1}.
      Please note, that the PHENIX data are for $0\cdots10\proz$
      centrality.
      The hatched band denotes our calculations for the 'Cronin'
      parameter $\alpha = 0.25\cdots0.4$ in (\ref{kt}).
      The solid line results from transport calculations
      without employing any initial state Cronin enhancement ($\alpha=0$).
      }
    \label{fig7}
  \end{center}
\end{figure}

We start with most central (5\% centrality) Au+Au collisions at
\SqrtS{200}. In this case  the nuclear modification factor is defined in
accordance with eq.~(\ref{ratio}) as
\begin{equation}
  \label{ratioAA}
  R_{\rm AA}(\pT) = \frac{1/N_{\rm AA}^{\rm event}\ d^2N_{\rm AA}/dy d\pT}
  {\left<N_{\rm coll}\right>/\sigma_{pp}^{\rm inelas}\ d^2
    \sigma_{pp}/dy d\pT}\ .
\end{equation}
Fig.~\ref{fig7} shows a comparison of the calculations for
eq.~(\ref{ratioAA}) with the data for charged hadrons from Ref.
\cite{PHENIX1,STAR1}, where the hatched band again
corresponds to the range in the parameter $\alpha$ in eq.~(\ref{kt})
from $0.25\dots0.4$.
The calculations roughly reproduce the shape of the ratio
$R_{\rm AA}(\pT)$ but overestimate the experimental ratio at higher $\pT$.
We emphasize, that the Cronin enhancement is visible
at all momenta, but does not show up to be responsible for the peak
structure in the enhancement around $2\GeVc$.
This becomes clear by comparing to the lower solid line, which results from
transport calculations without employing any initial state Cronin
enhancement. In this case the ratio
$R_{\rm AA}(\pT)$ is slightly low at $\pT\approx 2\dots3\GeVc$
in comparison to the data from the STAR collaboration, but rather well
in line with the data from PHENIX.
For $\pT>5\GeVc$, however, it still underestimates the suppression
seen by both collaborations.

\begin{figure}[htb!]
  \begin{center}
    \includegraphics[width=14.0cm]{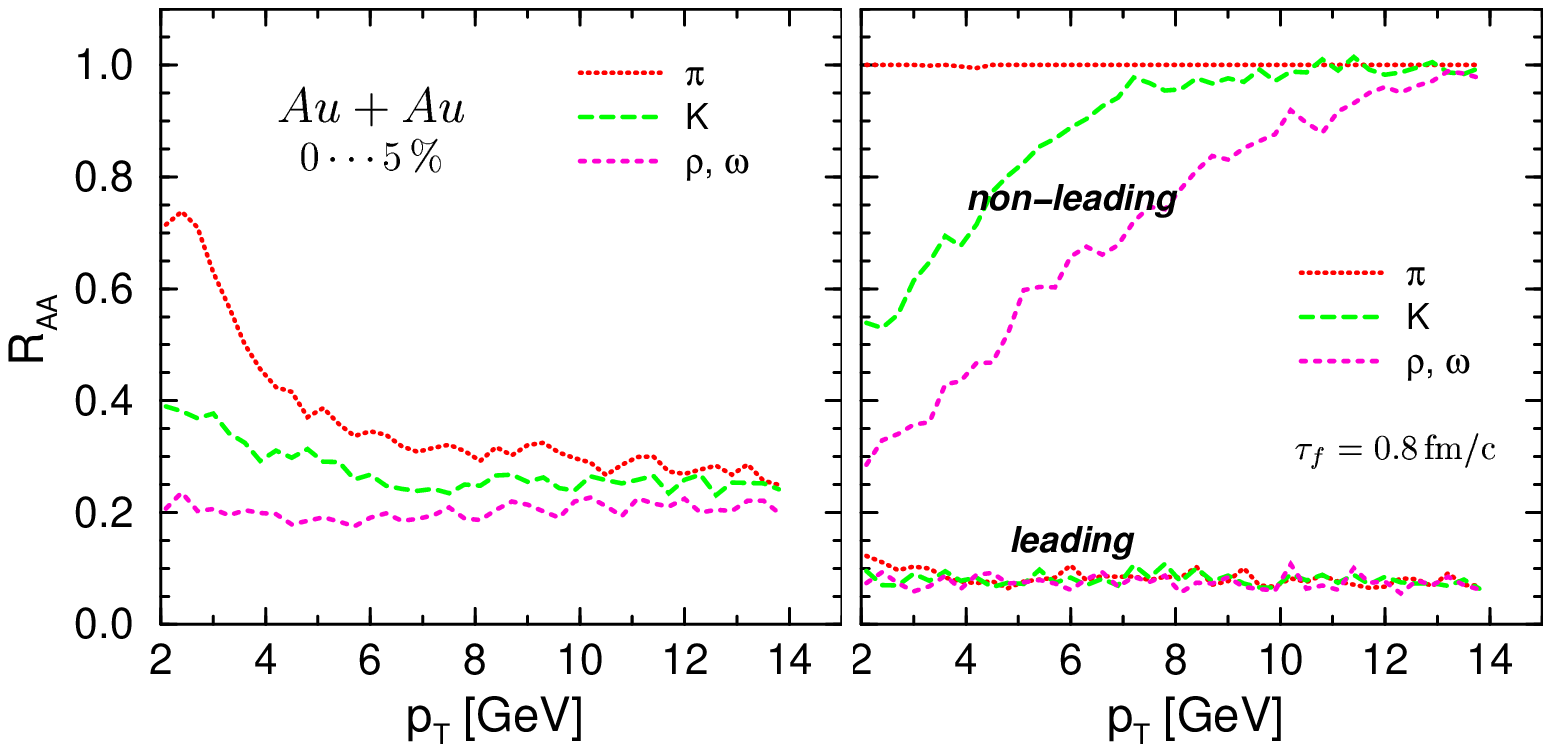}

    \caption{
      The nuclear modification factor for central Au+Au collisions
      (\SqrtS{200}) prior to any particle decays.
      The l.h.s.~shows the total theoretical ratios for the
      particle classes $\pi$, $\rho$, $\omega$ and $K$, $K^*$.
      The right panel separates the theoretical contributions of the
      left panel into the 'leading' and the 'non--leading' particle
      contributions.
      }
    \label{fig8}
  \end{center}
\end{figure}

In order to understand this result we decompose the ratio $R_{\rm AA}$
into contributions from pions, kaons and $\rho$, $\omega$, $K^*$
vector mesons (before decays) without employing any initial state
Cronin enhancement. The resulting comparison is shown in the left part
of Fig.~\ref{fig8} for 5\proz{} central Au+Au collisions.
Here the direct pions show a reduced attenuation, the kaon
reduction is slightly larger for lower $\pT$,
 while the vector meson absorption is much stronger.
Hadron formation time effects apparently play a substantial role
in the few $\GeVc$ region since heavier hadrons are formed earlier
than light pions in the cms frame at fixed transverse momentum due
to the lower Lorentz boost.

This expectation is quantified in the right part of
Fig.~\ref{fig8} where the decomposition of the ratio $R_{\rm AA}$ is
additionally performed for leading and secondary/nonleading
hadrons. First of all, the attenuation of the leading pions, kaons
and vector mesons is roughly independent on $\pT$ and hadron
type and gives $R_{\rm AA} \approx 0.09$. Thus the dominant
attenuation seen experimentally should be addressed to the
interactions of leading (unformed) pre-hadrons.
This becomes even more apparent in accordance with all
``back--of--the--envelope'' estimates (including a constant rest frame
formation time $\tau_f$) when looking at the suppression factors for the
non-leading hadrons in Fig.~\ref{fig8} (right part). Here the direct
pions practically do not show any attenuation in line with the
expectation from the large formation times of pions with momenta
of a couple of $\GeVc$. The situation changes for kaons and especially
vector mesons due to their larger mass. Now formed hadrons of a few
$\GeVc$ may interact with other formed hadrons and look 'thermalized',
i.e.~their final momenta show up in the soft part of the spectrum
(below $\pT\approx 2\GeVc$). Since the expanding
fireball has the same geometrical shape for all high $\pT$
hadrons, the attenuation in this ``constant $\tau_f$'' picture is
intimately correlated with the Lorentz $\gamma$--factor of the
particles such that for high $\pT$ mesons above about $6\dots8\GeVc$
no absorption of hadrons should occur via interactions of formed
hadrons. On the other hand, the final attenuation of high $\pT$
hadrons is approximately independent on the actual formation time
because according to Fig.~\ref{fig3} these particles are essentially
leading pre-hadrons that rescatter anyhow.

\begin{figure}[htb!]
  \begin{center}
    \includegraphics[width=7.67cm]{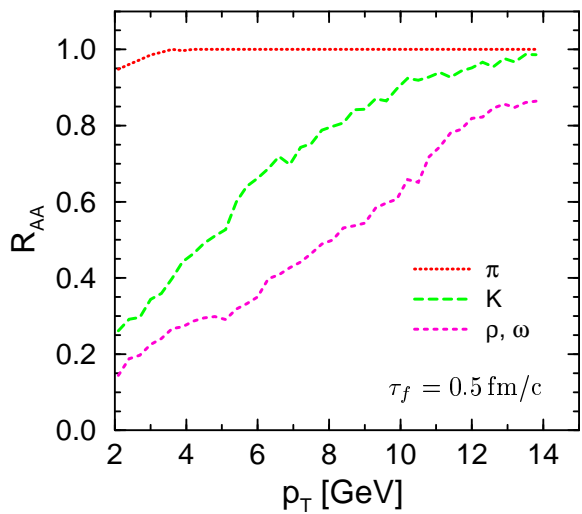}

    \caption{
      The same as Fig.~\ref{fig8} (right panel) for a
      formation time $\tau_f=0.5\fmc$. (Only the ratios of
      non-leading hadrons are displayed.)
      }
    \label{fig9}
  \end{center}
\end{figure}

In order to demonstrate the effect of the formation time $\tau_f$
we show in Fig.~\ref{fig9} our results for the attenuation of
secondary pions, kaons and vector mesons for $\tau_f=0.5\fmc$.
The latter value for $\tau_f$ is even favored by studies on
antiproton production and reabsorption in p+A reactions at AGS
energies \cite{AGS02} as well as hadron attenuation in virtual
photoproduction off nuclei \cite{Falter}. In line with the
discussion above we see from Fig.~\ref{fig9} -- in comparison to
the right part of Fig.~\ref{fig8} -- that the suppression of kaons
and vector mesons increases for a lower formation time and also
extends to higher $\pT$. However, as noted above, the total
attenuation is essentially determined by the fraction of leading
pre-hadrons at fixed $\pT$ (Fig.~\ref{fig3}) and by their
effective cross section with baryons or mesons. Formation time
effects can only show up at lower $\pT$.

We note in passing that a decrease in the energy density cut from
1 GeV/fm$^3$ to 0.7 GeV/fm$^3$ in the HSD calculations with
respect to the formation of secondary hadrons leads only to a
minor modification (enhancement) of the ratio $R_{AA}(\pT)$ by
about 10\%, i.e.~for $ \pT \geq6\GeVc$ the ratio increases from
$\sim$ 0.4 to $\sim$ 0.45.

\begin{figure}[htb!]
  \begin{center}
    \includegraphics[width=8cm]{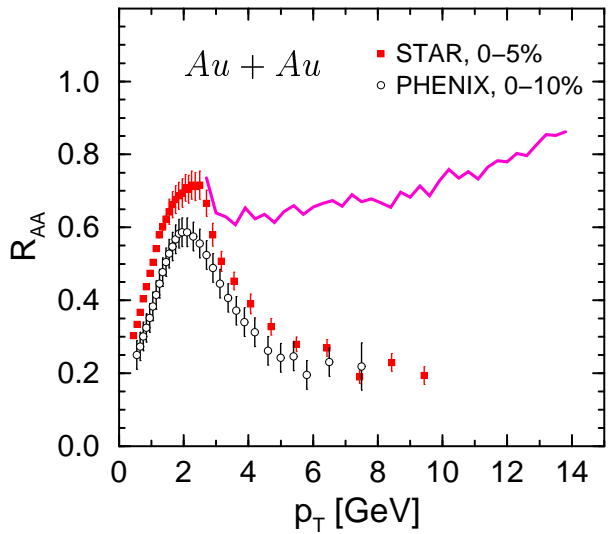}

    \caption{
      Same as fig.~\ref{fig7}, but with a leading cross section
      according to eq.~(\ref{sigma}) for the perturbative high $\pT$
      particles.
      }
    \label{fig10}
  \end{center}
\end{figure}

We now turn to an alternative model for the leading pre-hadron cross
section, which is discussed in the literature \cite{Dok} (cf.~also
Ref.~\cite{Kai}) 
since the notion of a fractional constant cross section might be
questionable \cite{Kopel4}
and alternative assumptions should be tested.\footnote{In the string
  fragmentation picture (combined with the constituent quark model)
  a considerable separation in space for the string ends at the starting
  time is unavoidable. The only possible pointlike structure here is
  the hard scattered quark from a string end.}
To this aim we have adopted a time--dependent cross section for
leading pre-hadrons of the kind
\begin{equation}
  \label{sigma}
  \sigma_{\rm lead}(\sqrt{s}) = \frac{t-t_0}{t_f} \sigma_{\rm had}
  (\sqrt{s})
\end{equation}
for $t-t_0 \leq t_f$, where $t_0$ denotes the actual production
time and $t_f$ its formation time in the calculational frame. The full
hadronic cross section is adopted for $t \geq t_0$. The numerical
results of this assumption are shown for the ratio $R_{\rm AA}$ in
Fig.~\ref{fig10} in case of 5\proz{} central Au+Au collisions.
We find that the concept eq.~(\ref{sigma}) is not in accordance
with the experimental observation since the leading pre-hadron cross
section is initially too low, i.e.~when the fireball is very compact
and dense. Furthermore, assuming eq.~(\ref{sigma}) in the
nonperturbative transport calculations leads to dramatic consequences:
In this case the pion, kaon and antibaryon rapidity distributions are
severely underestimated due to the lack of inelastic reactions from
leading pre-hadrons during the passage time of the heavy nuclei.

We thus continue our analysis with the 'default' formation time
$\tau_f = 0.8\fmc$ and leading pre-hadron cross sections in line with the
constituent quark model without attempting to 'fine--tune' any parameter.

\subsection{Dependence on centrality}

\begin{figure}[htb!]
  \begin{center}
    \includegraphics[width=14cm]{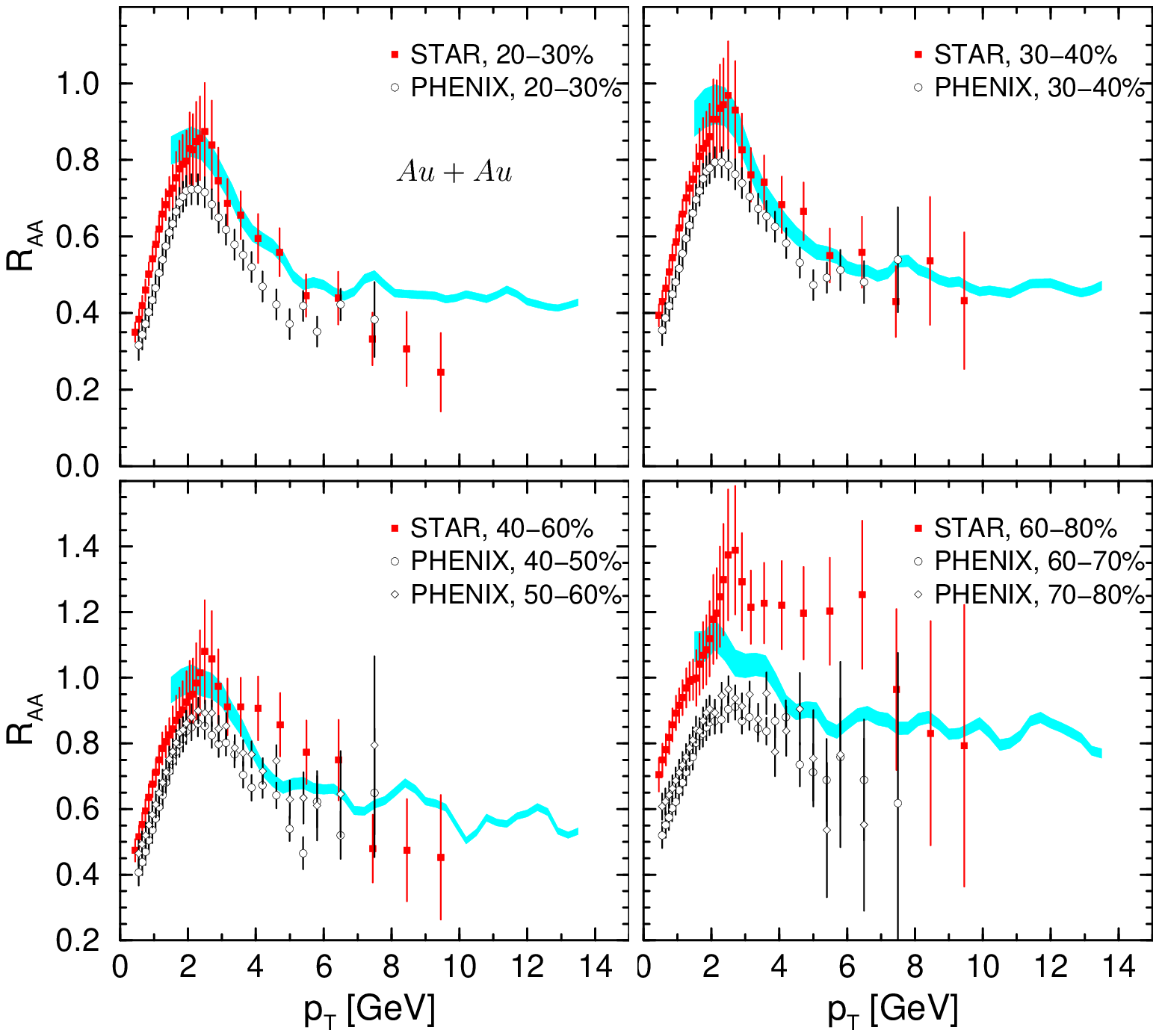}
    \caption{
      Same as Fig.~\ref{fig7} for the centralities
      $20\cdots30\proz{}$,  $30\cdots40\proz{}$, $40\cdots60\proz{}$ and
      $60\cdots80\proz{}$.
      Calculational fluctuations are due to limited statistics.
      Experimental data are from
      Refs.~\protect\cite{STAR1,PHENIX1}.}
    \label{fig11}
  \end{center}
\end{figure}

The centrality dependence of the ratio $R_{\rm AA}$ is shown in
Fig.~\ref{fig11} for
$20\cdots30\proz{}$, $30\cdots40\proz{}$, 
$40\cdots60\proz{}$ and $60\cdots80\proz{}$
centrality of Au+Au collisions at $\sqrt s=200\GeV$.
Again the hatched bands correspond to our calculations with a Cronin
parameter $\alpha$ in (\ref{kt}) ranging from 0.25 to 0.4 while the
data stem from Refs.~\cite{STAR1,PHENIX1}.
Note, that the uncertainty in the Cronin enhancement (width of the
hatched band) decreases for more peripheral reactions, which is a
direct consequence of the lower number of hard $NN$ collisions in
(\ref{kt}).
Our calculations turn out to be in a better agreement with the data
-- or interpolate between the measurements from the STAR and PHENIX
collaborations for 60 to 80\% centrality -- than for the most
central Au+Au collisions discussed in the previous subsection. An
analysis with respect to various contributions as in Fig.~\ref{fig7}
shows, that with decreasing centrality the influence of formed
hadrons essentially vanishes in the ratio $R_{\rm AA}(\pT)$ and the
final state suppression is dominated by the interaction of leading
pre-hadrons either 'immediately' with the baryons or other leading
debries of the colliding nuclei or with the hadrons formed at later
times.

\begin{figure}[htb!]
  \begin{center}
    \includegraphics[width=8.0cm]{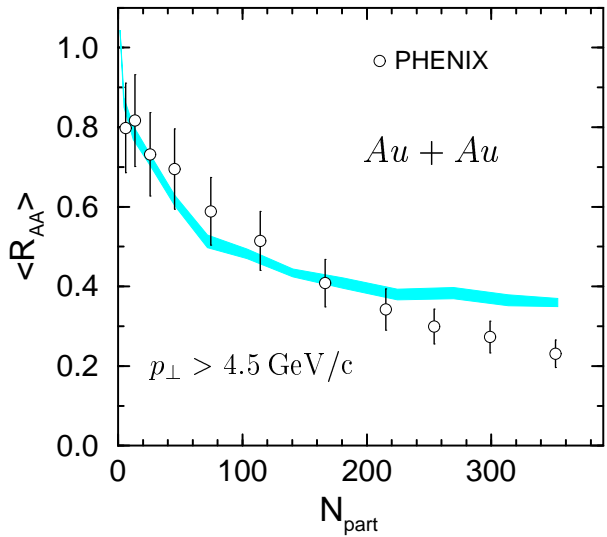}

    \caption{
      The ratio $\langle R_{AA}\rangle$ for charged hadrons for
      $\pT \geq 4.5\GeVc$ as a function of $N_{\rm part}$ in
      comparison to the data from Ref.~\cite{PHENIX1}.
      }
    \label{fig12}
  \end{center}
\end{figure}

In order to obtain a measure for the centrality dependence of the
suppression we display in Fig.~\ref{fig12} the suppression factor
$\langle R_{\rm AA}\rangle$ for charged hadrons integrated for
$\pT \geq 4.5\GeVc$
in comparison to the same quantity from the PHENIX collaboration
\cite{PHENIX1} as a function of the number of participating nucleons
$N_{\rm part}$.
We mention that the determination of $N_{\rm part}$ is model dependent
for very peripheral reactions, but rather save for mid--central and
central reactions. As already seen from Figs.~\ref{fig10} we
describe the average suppression factors well for
$N_{\rm part}\leq150$
and underestimate the suppression for central Au+Au interactions.

It is worth pointing out that the effective cross section of the
leading pre-hadrons is constant during their propagation and thus the
suppression is linear in the effective propagation length $L$. As
shown in a geometrical model by Drees et al.~\cite{Drees} this linear
absorption mechanism -- when fixed in strength to the most central
collisions -- overestimates the amount of suppression for
mid--central reactions. According to Ref.~\cite{Drees} the absorption
geometry favors a suppression $\sim L^2$ in comparison to the data in
Fig.~\ref{fig12}.
Though we do not find a strong argument in favor of any $\sim L^2$
suppression mechanism for all centrality classes we, nevertheless, seem
to need some additional attenuation for central Au+Au reactions which
could be attributed to strong interactions of the leading
pre-hadrons in a 'colored' medium.

\subsection{Elliptic flow and Angular Correlations}

Since the leading pre-hadron cross section employed in our transport
calculation is much larger than related cross sections from pQCD,
shadowing phenomena, that are related to the initial geometry
in coordinate space for fixed impact parameter $b$, should show up
in angular correlations like the elliptic flow of the high $\pT$
hadrons.
In fact, a rather high and approximately constant elliptic flow has
been seen experimentally for $\pT\geq2\GeVc$.
It is still to be answered, whether this elliptic flow is a
consequence of suppression due to strongly interacting matter or a
survivor of correlations from the hard nucleon--nucleon interactions
and following fragmentations.

Gating on particles with momenta $\pT\geq4\GeVc$ in the transport
calculation, only 1/3 of these final hadrons have suffered one or more
interactions during their propagation to the vacuum, whereas the other 2/3
escape without any interaction.
This observation implies, since more than 3/4 of the high $\pT$
hadrons are strongly absorbed, that the final high $p_T$ hadrons seen
experimentally essentially stem from pre--hadrons that originate from
a diffuse 'surface region' of the expanding fireball.
The latter pre--hadrons then evolve (or fragment) to the final hadrons
dominantly in the vacuum and are accompanied by secondary hadrons,
which - due to their large formation time - also hadronize in the
vacuum. Thus, when gating on high $\pT$ hadrons (in the vacuum) the
'near--side' correlations should be close to the 'near--side'
correlations observed for jet fragmentation in the vacuum. According
to our understanding the absence of 'near--side' jet broadening does
not necessarily indicate the presence of radiative quark energy loss
in a QGP phase as advocated in Ref. \cite{Wang03}.

We emphasize, that within our calculations the particles with large
transverse momenta -- that are treated perturbatively -- so far have
no intrinsic correlations as characteristic for back-to-back jet correlations.
This point will be addressed in the near future within a separate study.

The elliptic flow defined by
\begin{equation} \label{vv}
v_2(\pT) = \left\langle\frac{p_x^2-p_y^2}{p_x^2+p_y^2}\right\rangle_{\pT}
\end{equation}
provides a measure for the coupling of degrees of freedom in
coordinate space with those in momentum space. In eq.~(\ref{vv})
the $y$-direction denotes in--plane motion whereas the
$x$--direction denotes out--of--plane propagation. At RHIC
energies a strong in--plane elliptic flow is seen experimentally
which for $\pT\leq1.5\GeVc$ is rather well described by
hydrodynamical calculations \cite{hydro}. However, hydrodynamical
calculations overestimate the approximately constant elliptic flow
for higher transverse momenta. This observation proves that the
medium created in Au+Au collisions is strongly interacting and the
cross section needed to generate such a high elliptic flow is much
larger than expected from standard kinetic pQCD \cite{Miklos}.

\begin{figure}[htb!]
  \begin{center}
    \includegraphics[width=8.0cm]{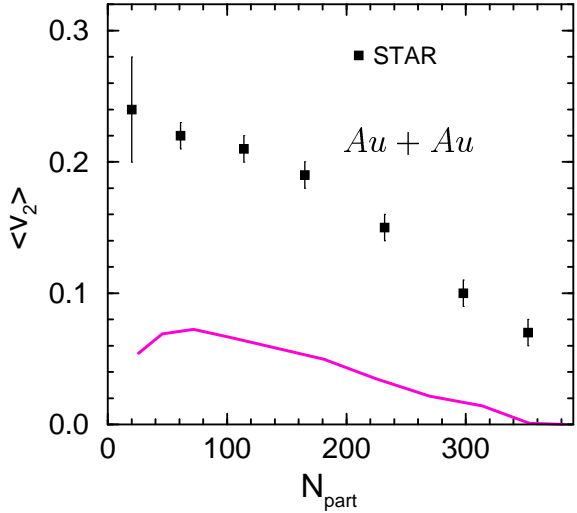}

    \caption{
      The elliptic flow $\langle v_2\rangle$ for charged hadrons for
      $\pT=2\cdots6\GeVc$ as a function of $N_{\rm part}$ in comparison
      to the data from Ref.~\cite{STARv2}.
      }
    \label{fig13}
  \end{center}
\end{figure}

In Fig.~\ref{fig13} we show the calculated elliptic flow
$\langle v_2\rangle $ for
charged hadrons with transverse momentum $\pT=2\cdots6\GeVc$ in
comparison to the data from Ref.~\cite{STARv2}.
One observes that our calculations underestimate the data by more than
a factor of 3; however, it is presently not clear if non--flow
contributions of different origin contribute to the measured $v_2$
values.
The elliptic flow described in our approach is essentially due to a
shadowing of high $\pT$ hadrons in the out--of--plane direction,
since there are no contributions from a collective pressure in the
perturbative dynamics of the high $\pT$ hadrons.
As already addressed by Shuryak \cite{Edward}, such shadowing effects
should be too weak to describe the large $v_2$ values seen experimentally.
The authors of Ref.~\cite{Drees} come to a very similar conclusion
in their geometrical model (cf. also Ref. \cite{Wang03}).

\subsection{SPS energies}

In this subsection we briefly comment on results obtained in
central collisions of Pb+Pb at \SqrtS{17.3}, i.e.~at top
SPS energies. We mention -- without explicit representation --
that the midrapidity $\pT$ spectra of hadrons from our \Pythia{}
calculations are approximately exponential in the transverse
momentum showing no explicit power--law shape as for
\SqrtS{200} in Fig.~\ref{fig1}.
This result stems in part from the kinematical cut in the maximum
transverse momentum, which is about $7\GeVc$ at \SqrtS{17.3} since two
strings (jets) have to be formed.
Furthermore, a direct comparison to related experimental spectra
is not possible because such data have not been measured. One might
extrapolate the hadron $\pT$ spectra from higher energies assuming a
$x_\perp$ scaling law to get a phenomenological parametrisation as in
Ref.~\cite{WA98}, however, our independent
attempt in this direction showed systematic, non--constant  errors (in
important parts of the spectra) up to a factor of 2 when employing
this ansatz.
This uncertainty has to be kept in mind especially when comparing
$\pT$ spectra from central Pb+Pb collisions to scaled 'extrapolated'
spectra from p+p reactions at the same $\sqrt{s}$.

\begin{figure}[htb!]
  \begin{center}
    \includegraphics[width=8.0cm]{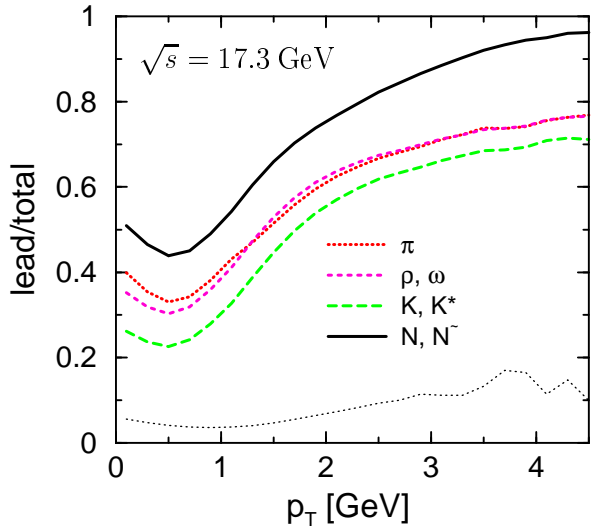}

    \caption{
      Same as Fig.~\ref{fig3} for \SqrtS{17.3}.
      (Note the different momentum scales.)
      The thin dotted line around 0.1 shows the contribution of
      leading baryons with 2 leading quarks.}
    \label{fig14}
  \end{center}
\end{figure}

Since at \SqrtS{17.3} the kinematical limits become essential for high
$\pT$ hadrons, the ratio of 'leading particles' to 'all produced
particles' (before particle decays) in N+N collisions changes
considerably in comparison to Fig.~\ref{fig3} for \SqrtS{200}.
This information is displayed in Fig.~\ref{fig14} for different
particle classes as a function of transverse momentum as calculated
within our \Pythia{} description.
A direct consequence of the results in Fig.~\ref{fig14} is that
above $\sim 2\dots3\GeVc$ now most of the mesons and the major
fraction of nucleons are of leading origin and -- within the dynamics
specified above -- may interact without delay.

\begin{figure}[htb!]
  \begin{center}
    \includegraphics[width=8cm]{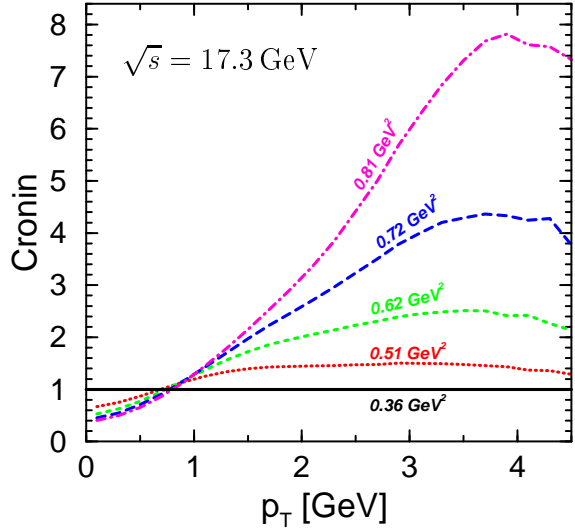}

    \caption{
      Same as Fig.~\ref{fig4} for \SqrtS{17.3}.
      (Note the different momentum scales and also the different sizes.)
      }
    \label{fig15}
  \end{center}
\end{figure}

A further consequence of the {\it steeper} transverse momentum spectra
at $\sqrt s=17.3\GeV$ relative to \SqrtS{200} (Fig.~\ref{fig1}) is that we
get a larger Cronin enhancement for the SPS energy.
This effect is due to a more drastic change in the spectra when
enhancing $\kTaveSq$ in the string fragmentation; as a consequence the
ratio between a calculation with enhanced $\kTaveSq$ to a calculation
with the default value for p+p collisions becomes larger as seen from
Fig.~\ref{fig15}.
This implies that the initial state Cronin enhancement is larger than
at RHIC energies (cf.~Fig.~\ref{fig4}).

\begin{figure}[htb!]
  \begin{center}
    \includegraphics[width=8.0cm]{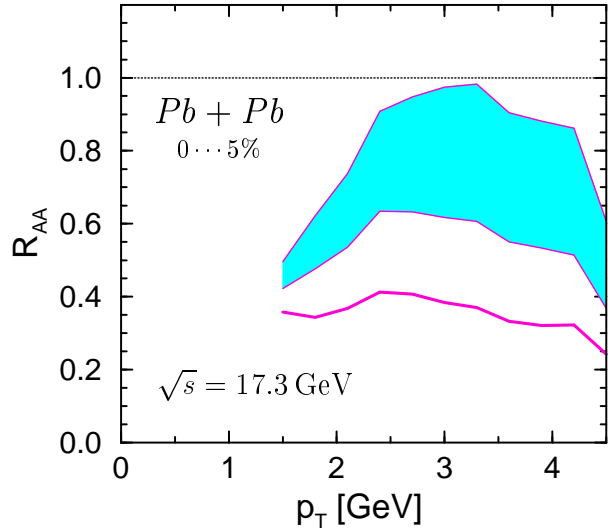}

    \caption{
      The suppression factor $R_{\rm AA}$ of charged hadrons
      at $0\cdots5\proz{}$ central $Pb+Pb$ collisions (\SqrtS{17.3})
      at midrapidity. The hatched band corresponds to our
      calculations with the parameter $\alpha$ in (\ref{kt}) ranging
      from 0.25 to 0.4. The solid line results from transport calculations
      without employing any initial state Cronin enhancement ($\alpha=0$).
      }
    \label{fig16}
  \end{center}
\end{figure}

Fig.~\ref{fig16} shows the result of our transport calculations for
the ratio (\ref{ratioAA}) in 5\proz{} central Pb+Pb collisions at
\SqrtS{17.3}, where the hatched band again corresponds to the range in
the parameter $\alpha$ in eq.~(\ref{kt}) from $0.25\cdots0.4$.
As in Fig.~\ref{fig6} the solid line shows a calculation without
any initial state Cronin enhancement. We see from
Fig.~\ref{fig16}, that the uncertainty in the ratio
(\ref{ratioAA}) due to missing constraints on the initial state
effects becomes large. The underlying physics, however, is the
same as at RHIC energies: the leading high $\pT$ pre-hadrons
interact with the baryons of the projectile/target as well as with
secondary formed hadrons. This leads to a net ratio of $\sim0.35$
for $\pT\geq2\GeVc$ when discarding any initial state enhancement
(solid line in Fig.~\ref{fig16}). However, this large attenuation
is counterbalanced by a strong initial state Cronin enhancement,
which leads to an increase of the ratio (\ref{ratioAA}) up to
$2.5\dots3\GeVc$ followed by a moderate decrease.

We have to mention that the WA98 collaboration has reported on the
ratio $R_{\rm AA}$ for neutral pions at top SPS energies
\cite{WA98} using extrapolations for the $\pi^0$ spectra from p+p
reactions taken from $\pi^0$ (and $\pi^\pm$) spectra at larger
energies. They quote a steep increase of $R_{\rm AA}$ with $\pT$
up to transverse momenta of $3.5\GeVc$ (although with large error
bars), which is not described by our transport calculations.
However, in view of the uncertainties in the normalization of the
ratio (and the ratio itself!) we have to wait for measurements at
RHIC for $\sqrt{s}\approx20\GeV$, where the transverse momentum
spectra from p+p, d+Au  and central Au+Au collisions can be
measured by the same detector setup at the same $\sqrt{s}$ in
order to clarify this issue from the experimental side.

\section{Summary}

Summarizing this study, we point out that (pre-) hadronic final state
interactions, as included by default in present transport approaches
like HSD, are able to approximately reproduce the high $\pT$
suppression effects observed in d+Au and Au+Au collisions at RHIC
(\SqrtS{200}).
This finding is remarkable, since the same dynamics also describe
the hadron formation and attenuation in deep--inelastic lepton
scattering off nuclei at HERMES \cite{Falter} quite well.
Additionally, this also holds for antiproton production and
attenuation in proton--nucleus collisions at AGS energies
\cite{AGS02}.

We have, furthermore, found that interactions of formed hadrons
after a formation time $t_f$ (which should be a matter of further
debate) are not able to explain the attenuation observed
experimentally for transverse momenta $\pT\geq6\GeVc$. However,
the shape of the ratio $R_{\rm AA}$ in transverse momentum $\pT$
reflects the presence of final state interactions of formed
hadrons in the $1\dots5\GeVc$ range \cite{Kai}.

In order to describe the experimental data below
$\pT\simeq5\GeVc$, a reasonable description of the initial state
``Cronin effect'' has to be taken into account, which has been modeled
in this work by increasing the average \kTaveSq{} of the string
fragmentation in subsequent hard collisions and has been fixed
approximately by the d+Au measurements at RHIC.
Such initial state effects are of pre--hadronic origin in a strongly
interacting medium. Since they also show up in energetic
proton--nucleus reactions \cite{Cronin1,Cronin2}, the initial state
effects cannot be considered as an indication for a phase transition
(or crossover) of the high density matter produced in Au+Au collisions
at relativistic energies.

We have demonstrated, that the centrality dependence of the
modification factor $R_{\rm AA}$ in Au+Au collisions at \SqrtS{200}
is well described for peripheral and mid--central
collisions on the basis of leading pre-hadron interactions.
However, the attenuation in central Au+Au collisions is
underestimated, which one may address to additional interactions of
partons in a colored medium that have not been accounted for in
our present transport studies.

We finally note, that the elliptic flow $v_2$ for high transverse
momentum particles is underestimated by at least a factor of 3
(cf. Fig.~\ref{fig13}).
This demonstrates that geometrical shadowing in out--of--plane
direction is not sufficient to get the right magnitude of the elliptic
flow seen experimentally, which supports the independent studies
in Refs. \cite{Edward,Drees}.

Since the high $\pT$ particles in the present work have no
intrinsic back-to-back jet-like correlations, we postpone any
further detailed statements to an upcoming investigation in the
near future. Moreover, further experimental studies on the
suppression of high momentum hadrons from d+Au and Au+Au
collisions down to $\sqrt{s}$ = 20 GeV will be necessary to
separate initial state Cronin effects from final state attenuation
and to disentangle to role of partons in a colored partonic medium
from those of pre-hadrons in a color-neutral hot and dense
fireball.

\vspace*{1mm}
The authors acknowledge valuable discussions with  E.~L.~Bratkovskaya,
D.{} d'Enterria,
Th.~Falter, B.~Jacak, C.~M.~Ko, and Z.~W.~Lin.


\end{document}